\documentclass[pra,aps,twocolumn,superscriptaddress]{revtex4-1}

\usepackage{amsmath,amsfonts,amssymb,caption,hyperref,color,epsfig,graphics,graphicx,latexsym,mathrsfs,revsymb,theorem,url,verbatim,epstopdf}
\usepackage{multirow} 

\usepackage{booktabs} 
\hypersetup{colorlinks,linkcolor={blue},citecolor={blue},urlcolor={red}}

\usepackage{hyperref}
\makeatletter

\newcommand{\Rmnum}[1]{\expandafter\@slowromancap\romannumeral #1@}
\makeatother
\newtheorem{definition}{Definition}
\newtheorem{proposition}[definition]{Proposition}
\newtheorem{Lemma}{Lemma}

\newtheorem{Theorem}{Theorem}
\newtheorem{Corollary}{Corollary}
\newtheorem{conjecture}[definition]{Conjecture}

\newtheorem{remark}{Remark}
\newtheorem{example}[definition]{Example}
\newtheorem{question}[definition]{Question}

\def\squareforqed{\hbox{\rlap{$\sqcap$}$\sqcup$}}
\def\qed{\ifmmode\squareforqed\else{\unskip\nobreak\hfil
		\penalty50\hskip1em\null\nobreak\hfil\squareforqed
		\parfillskip=0pt\finalhyphendemerits=0\endgraf}\fi}
\def\endenv{\ifmmode\;\else{\unskip\nobreak\hfil
		\penalty50\hskip1em\null\nobreak\hfil\;
		\parfillskip=0pt\finalhyphendemerits=0\endgraf}\fi}
\def\Dbar{\leavevmode\lower.6ex\hbox to 0pt
	{\hskip-.23ex\accent"16\hss}D}
\makeatletter
\def\url@leostyle{%
	\@ifundefined{selectfont}{\def\UrlFont{\sf}}{\def\UrlFont{\small\ttfamily}}}
\makeatother
\def\bcj{\begin{conjecture}}
	\def\ecj{\end{conjecture}}
\def\bcr{\begin{corollary}}
	\def\ecr{\end{corollary}}
\def\bd{\begin{definition}}
	\def\ed{\end{definition}}
\def\bea{\begin{eqnarray}}
	\def\eea{\end{eqnarray}}
\def\bem{\begin{enumerate}}
	\def\eem{\end{enumerate}}
\def\bex{\begin{example}}
	\def\eex{\end{example}}
\def\bim{\begin{itemize}}
	\def\eim{\end{itemize}}
\def\bl{\begin{lemma}}
	\def\el{\end{lemma}}
\def\bma{\begin{bmatrix}}
	\def\ema{\end{bmatrix}}
\def\bpf{\begin{proof}}
	\def\epf{\end{proof}}
\def\bpp{\begin{proposition}}
	\def\epp{\end{proposition}}
\def\bqu{\begin{question}}
	\def\equ{\end{question}}
\def\br{\begin{remark}}
	\def\er{\end{remark}}
\def\bt{\begin{theorem}}
	\def\et{\end{theorem}}

\def\btb{\begin{tabular}}
	\def\etb{\end{tabular}}

\newcommand{\nc}{\newcommand}

\nc{\bbA}{\mathbb{A}} \nc{\bbB}{\mathbb{B}} \nc{\bbC}{\mathbb{C}}
\nc{\bbD}{\mathbb{D}} \nc{\bbE}{\mathbb{E}} \nc{\bbF}{\mathbb{F}}
\nc{\bbG}{\mathbb{G}} \nc{\bbH}{\mathbb{H}} \nc{\bbI}{\mathbb{I}}
\nc{\bbJ}{\mathbb{J}} \nc{\bbK}{\mathbb{K}} \nc{\bbL}{\mathbb{L}}
\nc{\bbM}{\mathbb{M}} \nc{\bbN}{\mathbb{N}} \nc{\bbO}{\mathbb{O}}
\nc{\bbP}{\mathbb{P}} \nc{\bbQ}{\mathbb{Q}} \nc{\bbR}{\mathbb{R}}
\nc{\bbS}{\mathbb{S}} \nc{\bbT}{\mathbb{T}} \nc{\bbU}{\mathbb{U}}
\nc{\bbV}{\mathbb{V}} \nc{\bbW}{\mathbb{W}} \nc{\bbX}{\mathbb{X}}
\nc{\bbZ}{\mathbb{Z}}

\nc{\bA}{{\bf A}} \nc{\bB}{{\bf B}} \nc{\bC}{{\bf C}}
\nc{\bD}{{\bf D}} \nc{\bE}{{\bf E}} \nc{\bF}{{\bf F}}
\nc{\bG}{{\bf G}} \nc{\bH}{{\bf H}} \nc{\bI}{{\bf I}}
\nc{\bJ}{{\bf J}} \nc{\bK}{{\bf K}} \nc{\bL}{{\bf L}}
\nc{\bM}{{\bf M}} \nc{\bN}{{\bf N}} \nc{\bO}{{\bf O}}
\nc{\bP}{{\bf P}} \nc{\bQ}{{\bf Q}} \nc{\bR}{{\bf R}}
\nc{\bS}{{\bf S}} \nc{\bT}{{\bf T}} \nc{\bU}{{\bf U}}
\nc{\bV}{{\bf V}} \nc{\bW}{{\bf W}} \nc{\bX}{{\bf X}}
\nc{\bZ}{{\bf Z}}

\nc{\cA}{{\cal A}} \nc{\cB}{{\cal B}} \nc{\cC}{{\cal C}}
\nc{\cD}{{\cal D}} \nc{\cE}{{\cal E}} \nc{\cF}{{\cal F}}
\nc{\cG}{{\cal G}} \nc{\cH}{{\cal H}} \nc{\cI}{{\cal I}}
\nc{\cJ}{{\cal J}} \nc{\cK}{{\cal K}} \nc{\cL}{{\cal L}}
\nc{\cM}{{\cal M}} \nc{\cN}{{\cal N}} \nc{\cO}{{\cal O}}
\nc{\cP}{{\cal P}} \nc{\cQ}{{\cal Q}} \nc{\cR}{{\cal R}}
\nc{\cS}{{\cal S}} \nc{\cT}{{\cal T}} \nc{\cU}{{\cal U}}
\nc{\cV}{{\cal V}} \nc{\cW}{{\cal W}} \nc{\cX}{{\cal X}}
\nc{\cZ}{{\cal Z}}

\nc{\hA}{{\hat{A}}} \nc{\hB}{{\hat{B}}} \nc{\hC}{{\hat{C}}}
\nc{\hD}{{\hat{D}}} \nc{\hE}{{\hat{E}}} \nc{\hF}{{\hat{F}}}
\nc{\hG}{{\hat{G}}} \nc{\hH}{{\hat{H}}} \nc{\hI}{{\hat{I}}}
\nc{\hJ}{{\hat{J}}} \nc{\hK}{{\hat{K}}} \nc{\hL}{{\hat{L}}}
\nc{\hM}{{\hat{M}}} \nc{\hN}{{\hat{N}}} \nc{\hO}{{\hat{O}}}
\nc{\hP}{{\hat{P}}} \nc{\hR}{{\hat{R}}} \nc{\hS}{{\hat{S}}}
\nc{\hT}{{\hat{T}}} \nc{\hU}{{\hat{U}}} \nc{\hV}{{\hat{V}}}
\nc{\hW}{{\hat{W}}} \nc{\hX}{{\hat{X}}} \nc{\hZ}{{\hat{Z}}}

\nc{\hn}{{\hat{n}}}

\def\max{\mathop{\rm max}}
\def\min{\mathop{\rm min}}

\newcommand{\bra}[1]{\langle#1|}
\newcommand{\ket}[1]{|#1\rangle}

\def\Dbar{\leavevmode\lower.6ex\hbox to 0pt
	{\hskip-.23ex\accent"16\hss}D}

\begin{document}
	\title{Schmidt-number robustness as a unified quantifier of high dimensional entanglement in Buscemi nonlocality}
	
	\author{Xian Shi}\email[]
	{shixian01@gmail.com}
	\affiliation{College of Information Science and Technology,
		Beijing University of Chemical Technology, Beijing 100029, China}

	%
	
	
	
	\date{\today}
	
	\pacs{03.65.Ud, 03.67.Mn}
	
	\begin{abstract}
High-dimensional entanglement, captured by the Schmidt number, underpins advantages in quantum information tasks, yet a unified resource-theoretic description across different Buscemi-type operational objects has been missing. Here we develop a convex framework that treats bipartite states, distributed measurements, and teleportation instruments generated from shared entanglement on equal footing. For a fixed Schmidt-number threshold 
$k$, we introduce robustness-based monotones for each class of objects and prove a quantitative collapse: the Schmidt-number robustness of a bipartite state coincides with the maximal robustness achievable by any distributed measurement or teleportation instrument derived from that state. Consequently, within Buscemi-type operational frameworks, these objects do not carry independent high-dimensional resources but are governed by a single robustness-based monotone. We further provide a direct operational interpretation by relating this unique quantifier to the optimal advantage in entanglement-assisted state discrimination games. Our results complete a unified resource-theoretic characterization of high-dimensional entanglement across states, measurements, and quantum devices.

	\end{abstract}
	\maketitle

\emph{Introduction.}
High-dimensional entanglement has attracted increasing attention in quantum information science due to its fundamental significance and practical advantages in quantum communication, computation, and cryptography \cite{advances2020}. Compared with qubit entanglement, high-dimensional entangled systems enable enhanced information capacity, improved noise tolerance, and stronger security against eavesdropping, thereby motivating extensive theoretical and experimental investigations \cite{cui2020high,vigliar2021error,hu2023progress,bakhshinezhad2024scalable,zahidy2024practical,pauwels2025classification}. As a consequence, the characterization and quantification of high-dimensional entanglement have become central problems in modern quantum information theory \cite{bavaresco2018measurements,wyderka2023construction,liu2023characterizing,morenosemi,liu2024bounding,shi2024families,tavakoli2024enhanced,duquantifying,liangdetecting,mukherjee2025measurement}.

From the perspective of resource theories, entanglement is viewed as a valuable nonclassical resource whose manipulation is constrained by physically motivated free operations \cite{horodecki2009quantum,plenio2014introduction}. Traditional approaches to entanglement quantification focus primarily on quantum states, leading to a variety of entanglement measures and witnesses tailored to specific operational tasks \cite{horodecki1997separability,vidal1999robustness,terhal2000schmidt,lewenstein2001characterization,spengler2012entanglement,sperling2013multipartite,chruscinski2014entanglement,chen2015general,graydon2016entanglement,ketterer2020entanglement,rastegin2023,shi2024entanglement}. However, recent developments have revealed that nonlocal resources extend beyond states alone, encompassing more general objects such as distributed measurements and quantum instruments \cite{buscemi2012all,branciard2013measurement,nawareg2015experimental,mallick2017witnessing,cavalcanti2017all,bowles2018device,abiuso2021measurement,lipka2021operational}. 

In particular, the framework of Buscemi nonlocality provides a fully quantum generalization of Bell nonlocality, in which quantum inputs are allowed and nonlocality can be demonstrated without classical assumptions on measurement devices \cite{buscemi2012all}. Within this framework, quantum states, distributed measurements, and teleportation instruments naturally arise as distinct operational manifestations of nonlocal resources \cite{lipka2021operational}. Despite their apparent differences, it is of interest to ask whether these objects admit a unified quantitative description of high-dimensional entanglement. Furthermore, recent results \cite{lipka2020,lipka2021operational} mainly addressed on the separabie-entangled boundary, and donot consider the hierarchical structure of entanglement dimensions. Particularly, it remains unknown whether there exists a consistent monotone that is naturally tailored to the constraints with Schmidt number less than $k$, and quantitatively coincides across different Buscemi-type objects, states, distributed measurements, and teleprotation instruments,rather than relating them qualitatively merely.

In this work, we answer this question by developing a convex resource-theoretic framework in which bipartite states, distributed measurements, and teleportation instruments generated by shared entanglement are treated on equal footing. For each class of objects, we introduce robustness-based quantifiers associated with a fixed Schmidt number threshold and establish their quantitative equivalence. This demonstrates that the Schmidt-number robustness serves as a unique monotone consistently characterizing the resource content across states, measurements, and quantum devices.
Finally, we show that this unified resource quantifier admits a direct operational interpretation via entanglement-assisted state discrimination tasks. Rather than introducing a new application, this result completes the resource-theoretic picture by linking the abstract equivalence established above to experimentally relevant operational advantages.

Our contributions are as follows.
(i) We formulate a unified convex resource-theoretic framework for three Buscemi-type operational objects generated from shared entanglement: bipartite states, distributed measurements, and teleportation instruments. For a fixed Schmidt-number threshold $k$, we define robustness-based quantifiers for each class and provide SDP characterizations together with standard properties (faithfulness, convexity, and monotonicity).

(ii) We prove a quantitative collapse theorem: for any bipartite state $\rho$ and fixed $k$, the Schmidt-number robustness of $\rho$ equals the maximal robustness achievable by (a) teleportation instruments and (b) distributed measurements generated from $\rho$ via local POVMs. As consequences, we obtain an explicit restriction to generalized Bell-type POVMs for the teleportation part and a free-object equivalence across the three classes.

(iii) We give a direct operational interpretation of this unique quantifier through entanglement-assisted state-discrimination games, showing that the maximal relative advantage over $k$-Schmidt-number states is exactly $1+R_e^k(\rho)$, and we also address how optimal strategies can be constructed from the SDP dual witnesses.

\emph{Preliminaries}
	Assume $\mathcal{H}$ is a Hilbert space with finite dimensions, and the dimension of any Hilbert space is $d$. A quantum state is positive semidefinite with trace 1. A quantum substate is positive semidefinite with trace less than 1. Let $\boldsymbol{D}_A$ be the set of quantum states acting on $\mathcal{H}_A$. A quantum channel $\Delta_{A\rightarrow B}$ is a completely positive and trace-preserving linear map from $\boldsymbol{D}_A$ to $\boldsymbol{D}_B$. A subchannel $\Lambda_{A\rightarrow B}$ is completely positive and trace-nonincreasing. Let {$\boldsymbol{C}^{'}_{A\rightarrow B}$ and $\boldsymbol{C}_{A\rightarrow B}$ be the set of subchannels from $A$ to $B$ and the set of quantum channels from $A$ to $B,$ respectively. Obviously, $\boldsymbol{C}^{'}_{A\rightarrow B}\supset \boldsymbol{C}_{A\rightarrow B}.$ A positive operator valued measurement (POVM) $\mathbb{M}=\{M_i|i=1,2,\cdots,k\}$ is a set of positive semidefinite operators with $k$ outcomes and $\sum_i M_i=\mathbb{I}.$
	
	Assume $\Lambda\in \boldsymbol{C}_{A\rightarrow B^{'}}$ is a subchannel, there exists a one to one correspondence between a bipartite substate $S^{B^{'}B}$ and a subchannel $\Lambda\in {\boldsymbol{C}^{'}}_{A\rightarrow B^{'}}$ by Choi-Jamiolkowski isomorphism \cite{CHOI1975285}, 
	\begin{align*}
		(\Lambda\otimes\mathbb{I})(\ket{\Omega}\bra{\Omega})=S,\hspace{4mm}
		\Lambda(\rho)=\mathrm{tr}_B[(\mathbb{I}\otimes\rho^T)S].
	\end{align*}
	Here $\ket{\Omega}=\frac{1}{\sqrt{d}}\sum_i\ket{ii}.$
	\vspace{2mm}
	
	Assume $\ket{\psi}^{AB}$ is a pure state in $\mathcal{H}^A\otimes\mathcal{H}^B$, there always exist orthonormal bases $\{\ket{{i}}^A\}$ and $\{\ket{{i}}^B\}$ of $\mathcal{H}^A$ and $\mathcal{H}^B$, respectively such that
	\begin{align*}
		\ket{\psi}^{AB}=\sum_{i=1}^k{\lambda_i}\ket{ii}^{AB},
	\end{align*}
	here $\lambda_i>0$ and $\sum_i \lambda_i^2=1.$ Here the nonzero number $k$ is the Schmidt number of $\ket{\psi}^{AB}$, that is, $SN(\ket{\psi}^{AB})=k$. The Schmidt number of a mixed state $\rho^{AB}$ is defined as follows \cite{terhal2000schmidt},
	\begin{align}
		SN(\rho^{AB})=\min_{\rho=\sum_ip_i\ket{\psi_i}^{AB}\bra{\psi_i}}\max_i SN(\ket{\psi_i}^{AB}),
	\end{align}
	where the minimization takes over all the decompositions $\{p_i,\ket{\psi_i}\}_i$ of $\rho^{AB}=\sum_ip_i\ket{\psi_i}^{AB}\bra{\psi_i}$. Here we denote $\mathcal{F}_{sk}$ as the set of bipartite states with Schmidt number less than $k,$ $i.$ $e.$,
	\begin{align*}
		\mathcal{F}_{sk}=\{\rho^{AB}|SN(\rho^{AB})\le k,\rho^{AB}\in \mathcal{D}(\mathcal{H}^{AB})\}.
	\end{align*}
	
The Schmidt-number robustness $\mathrm{R}_{ke}(\cdot)$ for the high-dimensional entanglement is defined as 
	
\begin{align}
	\mathrm{R}_{ke}(\rho_{AB})=&\min\hspace{5mm} r\label{r0}\\
	\rho_{AB}+r\sigma_{AB}=&(1+r)\gamma_{AB}\nonumber\\
	\hspace{5mm}\sigma_{AB}\in \mathcal{D}(\mathcal{H}_{AB})&,\hspace{5mm} \gamma_{AB}\in \mathcal{F}_{sk}.\nonumber
\end{align}

\emph{High Dimensional Distributed Measurements.} 	Assume Alice and Bob share a mixed state $\rho_{A^{'}B^{'}}$, here we allow Alice and Bob to apply any bipartite POVMs $\mathbb{M}^{AA^{'}}$ and $\mathbb{M}^{BB^{'}}$, respectively. Alice and Bob are able to store and share classical information and quantum information. Let us denote $\mathbb{M}^{AB}=\{M^{AB}_{ab}\}$ as the POVMs generated under the following, 
	\begin{align}
		M^{AB}_{ab}=\mathrm{tr}_{A^{'}B^{'}}[(M_a^{AA^{'}}\otimes M_b^{BB^{'}})(\mathbb{I}^A\otimes\rho^{A^{'}B^{'}}\otimes\mathbb{I}^B)]. \label{pm}
	\end{align}
	
In this manuscript, we maintain the denotation of the POVMs generated above as \emph{distributed measurements}, and let $\mathcal{R}_{DM}$ denote the set of all distributed measurements. One can check that the set $\mathcal{R}_{DM}$ is convex. Next when the distributed measurements are generated from bipartite states $\rho^{A^{'}B^{'}}$ with Schmidt number less than $k$, we denote them as free POVMs, and the set of free POVMs is denoted as $\mathcal{F}_{kDM}$. Specifically, when $\rho^{A^{'}B^{'}}$ in (\ref{pm}) is a state with $SN(\rho^{A^{'}B^{'}})\le k$, $\rho^{A^{'}B^{'}}$ can be written as 
\begin{align*}
	\rho^{A^{'}B^{'}}=\sum_{\lambda}p_{\lambda}\rho^{A^{'}B^{'}}_{\lambda},
\end{align*}
here $p_{\lambda}>0$, $\sum_{\lambda}p_{\lambda}=1$, and $SN(\rho^{A^{'}B^{'}}_{\lambda})\le k$, $\forall \lambda$.
Hence, the associated distributed measurement takes the form 
\begin{align}
	M^{AB}_{ab}=\sum_{\lambda}p_{\lambda} M^{AB}_{ab|\lambda},\label{pm0}
\end{align}
where $M^{AB}_{ab|\lambda}=\mathrm{tr}[(M^{AA^{'}}_a\otimes M^{B^{'}B}_b)(\mathbb{I}^A\otimes \rho^{A^{'}B^{'}}_{\lambda}\otimes\mathbb{I}^B)],$ that is, any element in $\mathcal{F}_{kDM}$ can be written as (\ref{pm0}). As $\mathcal{F}_{sk}$ is convex, $\mathcal{F}_{kDM}$ is convex. That is, we define a convex quantum resource theory of measurements.

Assume $\mathbb{M}\in \mathcal{R}_{DM}$ is a distributed measurement, a problem is how to quantify $\mathbb{M}$ in this resource theory?  Here we address the problem by proposing a quantifier through the robustness-based method.

\begin{definition}\label{d1}
	Assume $\mathbb{M}=\{M_{ab}\}$ is a distributed measure, the quantifier based on the robustness method for this resource theory is defined as follows,
	\begin{subequations}
	\begin{align}
		\mathrm{R}_{kDM}(\mathbb{M})=&\min r\label{r1}\\
		\textit{s. t.}\hspace{3mm}& M_{ab}\le  (1+r)O_{ab}, \\
		&\{O_{ab}\}\in \mathcal{F}_{kDM} \hspace{4mm}\forall a,b.
	\end{align}
\end{subequations}
\end{definition}

The dual program of $\mathrm{R}_{kDM}$ in Definition \ref{d1} is placed in (\ref{rr2}), with this dual program, we present and prove the properties of $\mathrm{R}_{kDM}(\cdot)$, which can declare the validness of $\mathrm{R}_{kDM}(\cdot)$ in Sec. \ref{app1}.

\vspace{3mm}

\emph{High Dimensional Teleportation Instruments.} Quantum teleportation is one of the most fascinating and meaningful applications in quantum theory \cite{bennett1993teleporting}. In the original quantum teleportation scheme, Alice and Bob share a maximally entangled state, and Charlie sends Alice an unknown state. Alice first performs a Bell-state measurement on the local bipartite systems; she then sends her shared state and the measurement result to Bob, who applies a unitary to his entangled state according to the received information. Finally, Bob obtains the state Alice received initially.
Furthermore, the teleportation scheme can also be used to certify nonlocality. In 2017, the authors in \cite{cavalcanti2017all} explored the non-classicality of teleportation schemes and presented a method to verify bipartite entanglement. The study in \cite{lipka2020} subsequently addressed the resource theory of the quantum teleportation scheme. First we introduce the concept of a quantum teleportation instrument.
\begin{definition}\label{d2}
	Assume $\rho_{AB}$ is a bipartite mixed state, a teleportation instrument $\mathsf{\Lambda}^{A\rightarrow B}=\{\Lambda_i|\mathcal{D}(\mathcal{H}_A)\rightarrow\mathcal{D}(\mathcal{H}_B)\}$ is a collection of subchannels 
	\begin{align}
		\Lambda_i(\varphi^A)=\mathrm{tr}_{AA^{'}}[(M_{i}^{AA^{'}}\otimes\mathbb{I}^B)(\varphi^A\otimes\rho^{A^{'}B})].\label{c}
	\end{align}
	Here $\mathbb{M}=\{M^{AA^{'}}_i\}$ is a POVM.
\end{definition}

Here we denote the set of all subchannels generated by $(\ref{c})$ as $\mathcal{R}_{sc}$. We say a teleportation instrument $\mathsf{\Lambda}$ is free when the Schmidt number of the shared state $\rho^{AB}$ which generates $\Lambda$ is less than $k,$ and the free set of teleportation instrument $\mathsf{\Lambda}$ is denoted as $\mathcal{F}_{sc}$. For a set of subchannels $\mathsf{\Lambda}$, we could quantify the resource under the robustness method,
\begin{align}
1+	\mathrm{R}_{sc}(\mathsf{\Lambda})=&\hspace{5mm} 1+r\label{r5}\\
\textit{s. t.}&\hspace{5mm} (1+r)\Gamma_i-{\Lambda}_i\ge_{pos} 0,\nonumber\\
&\hspace{5mm}\mathsf{\Gamma}=\{\Gamma_i\}\in \mathcal{F}_{sc}.\nonumber
\end{align} 
Here $(1+r)\Gamma_i-{\Lambda}_i\ge_{pos}0$ means 
$	(1+r)\Gamma_i-\Lambda_i$ is completely positive.

\begin{remark}
In \cite{Designolle2021genuine}, the authors addressed the genuine high-dimensional quantum steering. They denote an assemblage $\{\sigma_{a|x}\}$ is $n$-preparable if $\sigma_{a|x}$  can be written as
		\begin{align*}
			\sigma^B_{a|x}=\mathrm{tr}_{A}[(A_{a|x}\otimes\mathbb{I})\rho^{AB}],
		\end{align*}
		here $\rho^{AB}$ is a mixed state with $SN(\rho^{AB})=n$, and $A_{a|x}$ are semidefinite positive with $\sum_a A_{a|x}=\mathbb{I},$ $\forall x.$ 
			Let $\Omega_n$ denote the set of all $n$-preparable assemblages. Due to the definition of $n$-preparable, there exists a hierarchical structure $\Omega_n\supset\Omega_{n-1}\supset\cdots\supset\Omega_1.$ 
	
In a teleportation instrument defined in Definition \ref{d2}, assume this task is to let Bob obtain a set of states $\{\omega_x\}$ Alice initially received. In the process of the scheme, the set $\{\omega_x\}$ are inputted into the set of subchannels $\mathsf{\Lambda}=\{\Lambda_i\}$, and an enssemble $\{\tau^B_{i|x}\}$ is obtained, here
\begin{align*}
\tau_{i|x}=\mathrm{tr}_{AA^{'}}[(M_i^{AA^{'}}\otimes\mathbb{I}^B)(\omega_x^A\otimes\rho^{A^{'}B})]. 
\end{align*}
As $SN(\omega_x^A\otimes\rho^{A^{'}B})=SN(\rho^{A^{'}B})$, the assemblage $\tau_{i|x}$ is $n$-preparable. 
\end{remark}

\emph{Relations between High Dimensional Distributed Measurements and Bipartite Mixed Entangled States.} 
Before presenting our main result, we emphasize that the objects considered in this work—bipartite quantum states, distributed measurements, and teleportation instruments—are fundamentally different in nature. They belong to distinct mathematical spaces and arise in different operational contexts. Consequently, there is no a priori reason to expect that a single entanglement monotone should consistently quantify the resource content across all these manifestations.
In the following, we show that this nontrivial consistency indeed holds when entanglement is quantified in terms of the Schmidt number. Specifically, we establish that the Schmidt-number robustness uniquely governs the resource content of bipartite states and all distributed measurements and teleportation instruments generated from them. This result identifies a common quantitative structure underlying these seemingly disparate objects and provides a unified resource-theoretic description of high-dimensional entanglement.

\begin{Theorem}\label{t2}
	Assume $\rho_{A'B'}$ is a bipartite state and fix $k$.
	Then
	\begin{align}
		R_{ke}(\rho_{A'B'})
		=&\max_{M_A} R_{\mathrm{sc}}\!\left(\Lambda^{A\to B}[\rho_{A'B'},M_A]\right)\nonumber\\
		=&\max_{M_A,M_B} R_{kDM}\!\left(M^{AB}[\rho_{A'B'},M_A,M_B]\right).\label{thf1}
	\end{align}
	Here the first maximization is over all POVMs $M_A=\{M^{AA'}_{a}\}$ used in
	the teleportation instrument defined in Eq.~(6), and the second maximization
	is over all local bipartite POVMs $M_A=\{M^{AA'}_{a}\}$ and
	$M_B=\{M^{B'B}_{b}\}$ used in the distributed measurement defined in Eq.~(\ref{pm}).

\end{Theorem}

\begin{figure*}[htbp] 
	\centering 
	\includegraphics[width=1.4\columnwidth]{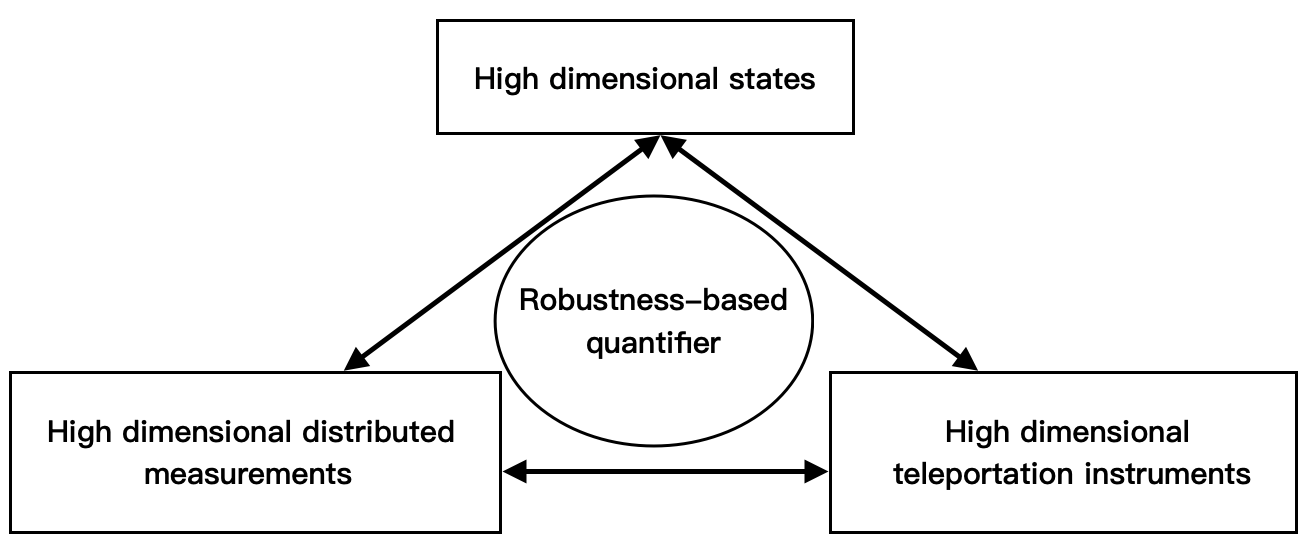} 
	\caption{Three Buscemi-type manifestations of high-dimensional entanglement: high-dimentional bipartite entanglement state, distributed measurements, and teleportation instruments. Theorem \ref{t2} shows that the resources can be completely characterized by the Schmidt-number robustness, yielding the quantitative equalities in (\ref{thf1}).} 
	\label{fig:label1} 
\end{figure*}
Here we place the proof of Theorem \ref{t2} in Sec. \ref{app2}. 
\begin{Corollary}
For the first equality in (\ref{thf1}), the maximization can be restricted 
	to generalized Bell-type POVMs of the form
	$M^{VA'}_{a}=(U^{V}_{a}\otimes I)(\psi^{+})$ with \emph{at most ${  d^{2} }$ outcomes,}
	where $\{U_a\}$ is a unitary operator basis (e.g. Pauli).
	An optimal choice can be constructed explicitly from an optimal dual witness
	$A^\star$ of the SDP for $R_e^{k}(\rho_{A'B'})$ (cf. Eq.~(\ref{r8})) via
	$Y^{VB}_{a}=\frac{1}{d}(U^{V}_{a}\otimes I)\left(A^{VB}\right)^{T}$.
\end{Corollary}
Theorem \ref{t2} tells us that the dimensions of distributed measurements and teleportation instruments generated by a bipartite state cannot be larger than $R_{ke}$.

\begin{Corollary}[Free-object equivalence]
	Fix $k$. For any bipartite state $\rho_{A'B'}$, the following statements are equivalent:
	\begin{enumerate}
		\item[(i)] $\rho_{A'B'}\in F_{S,k}$ (equivalently, $\mathrm{SN}(\rho_{A'B'})\le k$).
		\item[(ii)] For \emph{every} POVM $M_A$ used to generate a teleportation instrument,
		the induced instrument $\Lambda^{A\to B}[\rho_{A'B'},M_A]\in F_{T,k}$.
		\item[(iii)] For \emph{every} pair of POVMs $(M_A,M_B)$ used to generate a distributed measurement,
		the induced measurement $M^{AB}[\rho_{A'B'},M_A,M_B]\in F_{DM,k}$.
	\end{enumerate}

\end{Corollary}

\begin{example}
	Here we consider the family of the isotropic states on $\mathcal{H}_d\otimes\mathcal{H}_d$,
	\begin{align*}
		\rho_{iso}=\frac{1-F}{d^2-1}(\mathbb{I}-{\Psi})+F\Psi,
	\end{align*}
	here $\Psi=\ket{\psi_{+}}\bra{\psi_{+}}$,
	$\ket{\psi_{+}}=\frac{1}{\sqrt{d}}\sum_{i=0}^{d-1}\ket{ii}$.For the dual problem of $\mathrm{R}_{ke}(\rho)$ in \ref{r8}, based on the symmetric property of $\rho_{iso}$,  we could choose $A=\frac{d}{k}\Psi$ when $F>\frac{k}{d},$ otherwise, $A=0.$ Then based on the proof of theorem, 
	\begin{align*}
		\mathrm{R}_{ke}(\rho_{iso})=&\max_{M_A} R_{\mathrm{sc}}\!\left(\Lambda^{A\to B}[\rho_{iso},M_A]\right)\nonumber\\
		=&\max_{M_A,M_B} R_{kDM}\!\left(M^{AB}[\rho_{iso},M_A,M_B]\right)\\=&\max\{\frac{d}{k}F-1,0\}.
	\end{align*}
	Fig, \ref{fig2} is on the function of $\mathrm{R}_{ke}(\rho_{iso})$ when $d=5.$ 
	\begin{figure}[htbp] 
		\centering 
		\includegraphics[width=0.5\textwidth]{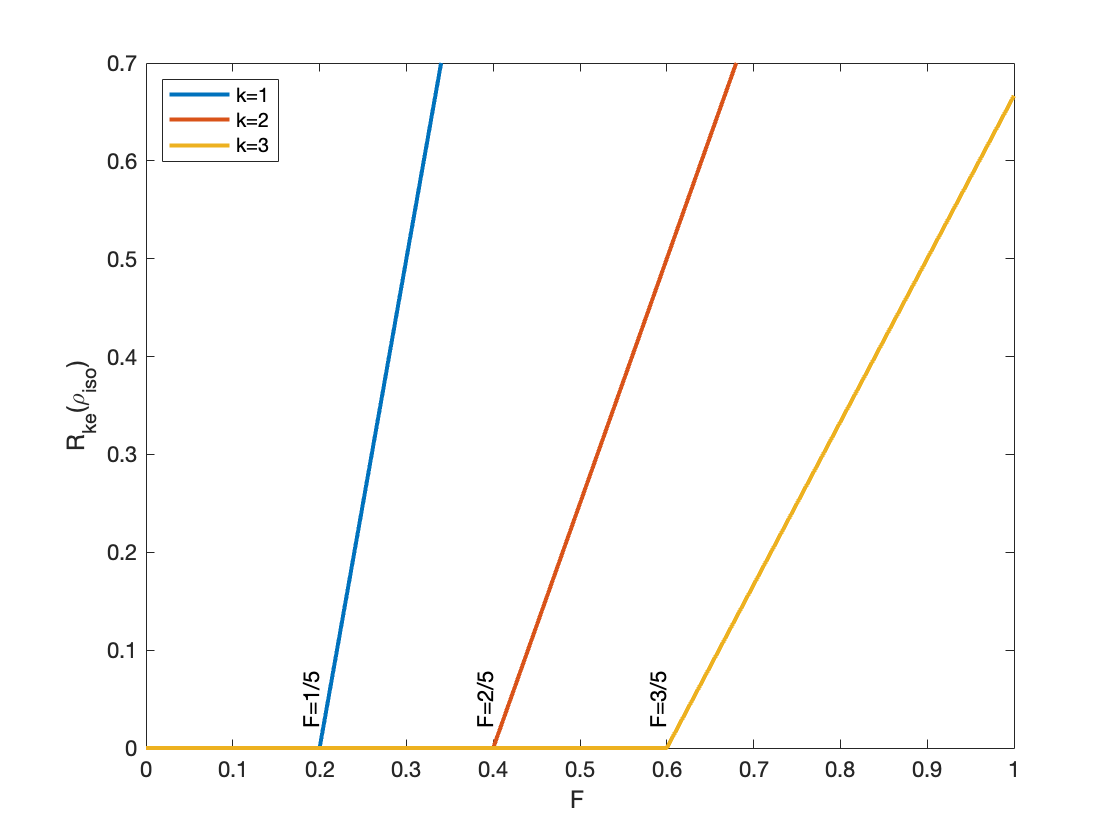} 
		\caption{The Schmidt-number robustness of $\rho_{iso}$.} 
		\label{fig2} 
	\end{figure}

\end{example}

In the next section, we present a task of quantum state discrimination to show the advantages of high-dimensional entanglement.

\emph{Entanglement-assisted state discrimination}.
Having established a unified robustness-based characterization of the Schmidt-number resource across states, distributed measurements, and teleportation instruments, it is natural to ask whether this abstract equivalence admits an operational interpretation. Importantly, the purpose of the following analysis is not to introduce a new discrimination protocol or a novel operational task. Instead, we use entanglement-assisted state discrimination as a canonical and well-understood scenario to demonstrate the operational completeness of the Schmidt-number robustness identified above.
In this sense, the discrimination game considered below should be viewed as a direct consequence of the resource-theoretic equivalence established in Theorem 1. It provides an explicit link between the unified resource quantifier and experimentally relevant performance metrics, thereby closing the conceptual gap between abstract resource measures and operational advantages.

Assume Charles owns a bipartite state from the ensemble $\mathcal{G}=\{p_{xy},\sigma_{xy}\}$ according to the probability distribution $\{p_{xy}\}$, then according to the prior probability distribution, he sends one part of the state to Alice, while sends the other part of the state to Bob. Next Alice and Bob choose arbitrary local POVMs $\mathbb{M}^A=\{M_{a}^{AA^{'}}\}$ and $\mathbb{M}^B=\{M_b^{BB^{'}}\}$ to the states they received and their part of the state $\rho^{A^{'}B^{'}}$ and receive outcomes $a$ and $b$, respectively. At last, Alice and Bob win the game only if they both guess which state they received. Hence, the average probability of the game can be expressed as
\begin{align*}
	p_{g}(\mathcal{G},\rho^{A^{'}B^{'}})=\max_{\mathbb{M}^A,\mathbb{M}^B}\sum_{a,b,x,y}p_{xy}\mathrm{tr}(M_{ab}\sigma_{xy})\delta_{xa}\delta_{yb},
\end{align*}
where the maximum takes over all the POVMs $\mathbb{M}^A$ of Alice and $\mathbb{M}^B$ of Bob with $M_{ab}$ ownning to the form like (\ref{pm}). When restricting the shared state $\rho^{A^{'}B^{'}}$ in $\mathcal{F}_{sk}$, the best average probability of the above task is denoted as
\begin{align*}
	p_{g}^{(k)}(\mathcal{G})=\max_{\delta^{A^{'}B^{'}}\in\mathcal{F}_{sk}}p_g(\mathcal{G},\delta^{A^{'}B^{'}}),
\end{align*}
where the maximum takes over all the bipartite states with its Schmidt number less than $k.$

\begin{proposition}\label{t5}
	Assume $\rho_{AB}$ is a bipartite state shared between Alice and Bob, let $\mathcal{G}=\{p_{xy},\sigma_{xy}\}$ be any ensemble of the bipartite states, then we have the following results,
	\begin{align*}
		\max_{\mathcal{G}}\frac{p_g(\mathcal{G},\rho^{AB})}{p_{g}^{(k)}(\mathcal{G})}=1+\mathrm{R}_{ke}(\rho^{AB}).
	\end{align*}
	
	Here the maximum takes over all the ensembles $\mathcal{G}=\{p_{xy},\sigma_{xy}\}$ of bipartite states.  
	
\end{proposition} 

Here we place the proof of Proposition \ref{t5} in Sec. \ref{app3}.

\emph{Conclusions} In this work, we investigated the certification and quantification of high-dimensional entanglement within the operational framework of Buscemi nonlocality. Rather than proposing a new operational task or resource model, our focus was on the structural content of resource theories constrained by a fixed Schmidt number.
Our main result establishes that, under such constraints, resource quantification collapses across different physical objects. Specifically, we showed that the Schmidt-number robustness of a bipartite state necessarily coincides with the maximal robustness achievable by any distributed measurement or teleportation instrument generated from that state. This implies that bipartite states, distributed measurements, and quantum devices do not carry independent high-dimensional resources within the Buscemi-type operational framework, but are instead governed by a single, unique robustness-based monotone.
An immediate consequence of this collapse is a no-refinement principle within the considered Buscemi-type frameworks and $k$-dimensional constraints:  whether based on distributed measurements, teleportation instruments, or discrimination games, can distinguish high-dimensional entanglement beyond its Schmidt-number robustness. In this sense, our results provide a classification and completion of Buscemi-type resource theories in the high-dimensional regime.
Finally, by connecting this unique quantifier to entanglement-assisted state discrimination games, we furnished a direct operational interpretation of the structural results, thereby closing the gap between abstract resource-theoretic analysis and experimentally relevant tasks.
Looking forward, it would be interesting to investigate whether similar collapse phenomena arise in multi-shot scenarios, as well as for other high-dimensional quantum resources beyond entanglement. Such extensions may further clarify the fundamental limits of operational resource theories in complex quantum systems

  \emph{Acknowledgement}
X. S. was supported by the National Natural Science Foundation of China (Grant No. 12301580).

\appendix
\section{Supplemental Materials}

In this section, we would first recall the knowledge needed of semidefinite programming (SDP), then we present the properties of $\mathrm{R}_{kDM}(\mathbb{M})$, next we consider the analytical expressions of $\mathrm{R}_{sc}(\Lambda)$ by a semidefinite programming. At last, we prove Theorem \ref{t2} in the main text.

Firstly, we recall the knowledge needed of SDP. Readers can refer to \cite{watrous2018theory,gour2019comparison} to learn more. Let $V_1$ and $V_2$ be two vector spaces, and let $K_1\subset V_1$ and $K_2\subset V_2$ be two convex cones. If $\Gamma:V_1\rightarrow V_2$ is a linear map, $H_1$ and $H_2$ are two given elements in $V_1$ and $V_2$, respectively,

The primal problem:
\begin{align}
	\alpha=&\hspace{5mm}\min \mathrm{tr}(XH_1)\label{p}\\
	\textit{s. t.}& \hspace{5mm}\Gamma(X)-H_2\in K_2,\nonumber\\
	&\hspace{5mm}X\in K_1.\nonumber
\end{align}
The dual problem:
\begin{align}
	\beta=&\hspace{5mm}\max\mathrm{tr}(YH_2)\label{d}\\
	\textit{s. t.}&\hspace{5mm} H_1-\Gamma^{*}(Y)\in K_1^{*},\nonumber\\
	&\hspace{5mm}Y\in K^{*}_2.\nonumber
\end{align}
Here $\Gamma^{*}:V_2\rightarrow V_1$ is the dual map of $\Gamma$, $K_1$ and $K_2$ are the dual cones of $K_1$ and $K_2$, respectively.

Next we present the strong duality conditions:
\begin{itemize}
	\item[1.] Assume $K\subset V_2\oplus \mathbb{R}$ is a cone defined as
	\begin{align*}
		K=\{(\Gamma(X)-Y,\mathrm{tr}XH_1):X\in K_1,Y\in K_2\}.
	\end{align*}
	If $K$ is closed in $V_2\oplus\mathbb{R}$ and there exists a primal feasible plan, $\alpha=\beta.$ 
	\item[2.] The Slater's condition: suppose that there is a primal feasible plan $X_0\in \emph{int}(K_1)$ such that $\Gamma(X_0)-H_2\in\emph{int}(K_2)$. If there exists a primal optimal plan, then $\alpha=\beta.$  
\end{itemize}

\subsection{Properties of $\mathrm{R}_{kDM}(\mathbb{M})$}\label{app1}
In this subsection, first we present the dual problem of $\mathrm{R}_{kDM}(\mathbb{M})$ for a POVM $\mathbb{M}$. Then we show the properties of $\mathrm{R}_{kDM}(\mathbb{M})$ defined in Theorem \ref{t1}.

Here we first present the dual problem of $\mathrm{R}_{kDM}(\mathbb{M}).$ Assume $\mathbb{M}$ is a POVM, then $\mathrm{R}_{kDM}(\mathbb{M})$ can be written as follows,

\begin{align}
1+	\mathrm{R}_{kDM}(\mathbb{M})=\hspace{5mm}&\min \frac{1}{d^2}\mathrm{tr}\sum_{ab}\tilde{O}_{ab}\label{r2}\\
\textit{s. t.}\hspace{5mm}& M_{ab}\le \tilde{O}_{ab}, \nonumber\\
&\{\tilde{O}_{ab}\}\in \tilde{\mathcal{F}}_{kDM}\hspace{4mm}\forall a,b.\nonumber
\end{align}

Here $\tilde{O}_{ab}=(1+r)O_{ab},$ and $\tilde{\mathcal{F}}_{kDM}$ is the cone of $\mathcal{F}_{kDM}.$ As $\mathbb{O}$ is a POVM, \begin{align*}
	\sum_{ab}O_{ab}=\mathbb{I}_{AB},\hspace{5mm}
	\mathrm{tr}\sum_{ab}O_{ab}=d^2,
\end{align*} here $dim(\mathcal{H}_A)=dim(\mathcal{H}_B)=d.$
The duality problem of (\ref{r2}) is 
\begin{align}
	1+\mathrm{R}_{kDM}(\mathbb{M})=\hspace{5mm}&\frac{1}{d^2}\max\sum_{ab}Y_{ab}M_{ab}\label{rr2}\\
	\textit{s. t.}\hspace{5mm}& \mathbb{I}-Y_{ab}\in \tilde{\mathcal{F}}_{kDM}^{*},\nonumber\\
&	Y_{ab}\ge0.
\end{align}

Moreover,  (\ref{r2}) can also be written as follows,

	\begin{align}
		1+	\mathrm{R}_{kDM}(\mathbb{M})=\hspace{5mm}&1+\min r\label{r3}\\
		\textit{s. t.}\hspace{5mm}&M_{ab}\le \tilde{O}_{ab}\nonumber\\
		&\sum_{a,b}\tilde{O}_{ab}=(1+r)\mathbb{I}^{AB}\nonumber\\
		&\tilde{\mathbb{O}}\in \tilde{\mathcal{F}}_{kDM}.\nonumber
			\end{align}

Then we present the duality problem of (\ref{r3}),
\begin{align}
	1+\mathrm{R}_{kDM}(\mathbb{M})=&\hspace{5mm}\max \sum_{a,b}\mathrm{tr}(Y_{ab}M_{ab})\nonumber\\
	\textit{s. t.}&\hspace{5mm} K-Y_{ab}\in \tilde{\mathcal{F}}_{kDM}^{*},\nonumber\\
	&\hspace{5mm}\mathrm{tr}K=1,\hspace{4mm}Y_{ab}\ge0.\label{r4}
\end{align}
Here $\tilde{\mathcal{F}}^{*}_{kDM}=\{Z|\mathrm{tr}(XZ)\ge 0,X\in \tilde{\mathcal{F}}_{kDM}\}$.

 Next we would recall the definition of simulability of distributed measurements, first we introduce the definition of $n$-\emph{partially entanglement breaking} ($n$-PEB). 
\begin{definition}\cite{Chruscinski2006on}
	A channel $\Lambda$ is $n$-PEB if $SN(\Lambda\otimes\mathbb{I}(\rho^{AB}))\le n$ for all bipartite states $\rho^{AB}.$
\end{definition}

The definition of simulability of distributed measurements in the resource theory considered here is based on $n$-PEB.

\begin{definition}
	Assume $\mathbb{M}^{AB}=\{M^{AB}_{ij}\}$ and $\mathbb{N}^{AB}=\{N^{AB}_{mn}\}$ are two distributed POVMs, we say $\mathbb{N}^{AB}$ can be simulated by $\mathbb{M}^{AB}$, if
	\begin{align}
		N^{AB}_{mn}=\sum_{ij\lambda}p_{\lambda}p(m,n|i,j,\lambda)\mathcal{E}_{\lambda}^{\dagger}[M^{AB}_{ij}],\label{npeb}
	\end{align}
	here $p_{\lambda}$ is a probability distribution, $\mathcal{E}_{\lambda}$ is $n$-PEB. When a POVM $\mathbb{N}$ can be simulated by a POVM $\mathbb{M}$ under the way defined in (\ref{npeb}), we denote $\mathbb{N}\preceq\mathbb{M}.$
\end{definition}

\begin{Theorem}\label{t1}
	Assume $\mathbb{M}^{AB}\in \mathcal{R}_{kDM}$ is a distributed measurement, the quantifier $\mathrm{R}_{kDM}(\mathbb{M})$ defined in Definition \ref{d1} satifies the following properties,
	\begin{itemize}
		\item[1.] [Faithfulness] $\mathrm{R}_{kDM}(\mathbb{M})$ vanishes if and only if the measurement $\mathbb{M}$ is free.
		\begin{align*}
			\mathrm{R}_{kDM}(\mathbb{M})=0\hspace{5mm} \Longleftrightarrow\hspace{5mm} \mathbb{M}\in \mathcal{F}_{kDM}.
		\end{align*} 
		\item[2.] [Convexity] Assume $\mathbb{M}_1$ and $\mathbb{M}_2$ are two distributed measurements in $\mathcal{R}_{DM},$ let  $\mathbb{M}=p\mathbb{M}_1+(1-p)\mathbb{M}_2$, 
		\begin{align*}
			\mathrm{R}_{kDM}(\mathbb{M})\le p\mathrm{R}_{kDM}(\mathbb{M}_1)+(1-p)\mathrm{R}_{kDM}(\mathbb{M}_2),
		\end{align*}
		here $p\in (0,1).$
		\item[3.] [Monotonicity] Assume $\mathbb{N}$ and $\mathbb{M}$ are two distributed measurement, if $\mathbb{N}$ can be simulated by $\mathbb{M}$ under the simulation strategy defined by (\ref{npeb}), $i.$ $e.$ $\mathbb{N}\preceq\mathbb{M},$ then
		\begin{align*}
			\mathrm{R}_{kDM}(\mathbb{N}^{AB})\le \mathrm{R}_{kDM}(\mathbb{M}^{AB}).
		\end{align*}
	\end{itemize}
\end{Theorem}

\emph{Proof of Theorem \ref{t1}:}

\begin{itemize}
	\item[1.] [Faithfulness] $\Longrightarrow:$ If $\mathbb{M}\in \mathcal{F}_{kDM}$, then we can choose $r=0$ in (\ref{r1}), as $\mathrm{R}_{kDM}(\cdot)$ is nonnegative,  $r=0$ is optimal.
	
	$\Longleftarrow:$ Due to the definition of $\mathrm{R}_{kDM}(\cdot)$, it is straightword to obtain that $\mathbb{M}\in\mathcal{F}_{kDM}$.
	
	\item[2.] [Convexity] Assume $\tilde{\mathbb{O}}_1$ and $\tilde{\mathbb{O}}_2$ are two optimal POVMs in $\tilde{\mathcal{F}}_{kDM}$ in terms of $\mathrm{R}_{kDM}(\cdot)$ for $\mathbb{M}_1$ and $\mathbb{M}_2$, respectively. As the set of bipartite states with Schmidt number less than $k$ is convex, $p\tilde{\mathbb{O}}_1+(1-p)\tilde{\mathbb{O}}_2\in\mathcal{F}_{kDM}.$ As 
	\begin{align*}
&1+\mathrm{R}_{kDM}(\mathbb{M}^{'})\\
\le& \frac{1}{d_Ad_B}[p\mathrm{tr}(\sum_{ab}\tilde{O}_{ab}^1)+(1-p)\mathrm{tr}(\sum_{ab}\tilde{O}_{ab}^2)]\\
=& p\mathrm{R}_{kDM}(\mathbb{M}_1)+(1-p)\mathrm{R}_{kDM}(\mathbb{M}_2)+1.
	\end{align*} 
	
	\item[3.] [Monotonicity] As $\mathbb{N}\preceq\mathbb{M},$ there exist a probability distribution $\{p_{\lambda}\}$ and $n$-PEB $\{\mathcal{E}_{\lambda}\}$ such that 
	\begin{align*}
		N_{mn}=\sum_{ij\lambda}p_{\lambda}p(m,n|i,j,\lambda)\mathcal{E}_{\lambda}^{\dagger}[M_{ij}].
	\end{align*} 
	
	Next assume $\mathbb{Y}=\{Y_{mn}\}$ and $K$ are the optimal for $\mathbb{N}$ in terms of the definition of $\mathrm{R}_{B}(\cdot)$, let $Z_{ij}=\sum p_{\lambda}p(a,b|i,j,\lambda)\mathcal{E}_{\lambda}[Y_{ab}]$. Let $\mathbb{H}=\{H_{mn}\}$ and $K$ be arbitrary in $\tilde{\mathcal{F}}_{kDM}$ and $\mathcal{F}_{sk}$, respectively, then
	\begin{align*}
	&\mathrm{tr}[H_{mn}(K-\sum_{i,j,\lambda}p_{\lambda}p(m,n|i,j,\lambda)\mathcal{E}_{\lambda}[Y_{mn}])]\\
	=&\mathrm{tr}[H_{mn}(\sum_{\lambda}\mathcal{E}_{\lambda}(K-\sum_{m,n}p_{\lambda}p(m,n|i,j,\lambda)Y_{mn}))]\\
	\ge&\mathrm{tr}[H_{mn}(\sum_{\lambda}\mathcal{E}_{\lambda}\sum_{m,n}p_{\lambda}p(m,n|i,j,\lambda)(K-Y_{mn}))]\\
	\ge&0
	\end{align*}
	 Here the first equality is due to that $\sum_{\lambda}\mathcal{E}_{\lambda}$ is a channel, and $\mathrm{tr}\sum_{\lambda}\mathcal{E}_{\lambda}(K)=\mathrm{tr}K.$ By the definition of $Y_{mn}$ and $(\ref{r4})$, the last inequality is valid. That is, $\mathbb{Z}=\{Z_{ab}\}$ satisfies the conditions of the dual problem (\ref{r4}). Hence,
	 \begin{align*}
	 	\mathrm{R}_{kDM}(\mathbb{M})\ge &\sum_{i,j}\mathrm{tr}(Z_{ij}M_{ij})\\
	 	=&\sum_{i,j}\mathrm{tr}(\sum p_{\lambda}p(a,b|i,j,\lambda)\mathcal{E}_{\lambda}[Y_{ab}]M_{ij})\\
	 	=&\sum_{a,b}\mathrm{tr}(\sum_{ij\lambda}p_{\lambda}p(a,b|i,j,\lambda)Y_{ab}\mathcal{E}_{\lambda}^{\dagger}(M_{ij}))\\
	 	=&\sum_{i,j}\mathrm{tr}(Y_{ij}N_{ij})\\
	 	=&\mathrm{R}_{kDM}(\mathbb{N}).
	 \end{align*}
	 Then we finish the proof.
\end{itemize}

\subsection{Properties of $\mathrm{R}_{sc}(\mathsf{\Lambda})$.  }
In this part, before presenting the properties of $\mathrm{R}_{sc}(\mathsf{\Lambda})$, we would restate $\mathrm{R}_{sc}(\cdot)$. Assume $\{M_{a}^{AA^{'}}\}$ is a POVM, $\rho_{A^{'}B}$ is a state shared by Alice and Bob. Next we let $J_a$ denotes the Choi-Jamiolkowski isomorphism of $\Lambda_a$, that is, $J_a=(\mathbb{I}\otimes \Lambda_a)[\ket{\psi_{+}}\bra{\psi_{+}}]$, here $\ket{\psi_{+}}=\frac{1}{\sqrt{d}}\sum_{i=1}^d\ket{ii}$ is the maximally entangled state. Hence, $\mathrm{R}_{sc}(\Lambda)$ can be written as 
\begin{align}
1+	\mathrm{R}_{sc}(\mathsf{\Lambda})=&\hspace{5mm}\min \mathrm{tr}\tilde{\sigma}^{B^{'}}\label{r6}\\
	\textit{s. t.}&\hspace{5mm} J_a^{VB^{'}}\le O_a^{VB^{'}} \hspace{3mm}\forall a,\nonumber\\
	&\hspace{5mm}\sum_a O_a^{VB^{'}}=\frac{\mathbb{I}}{d}\otimes\tilde{\sigma}^{B^{'}},\nonumber\\
&\hspace{5mm}	O_a^{VB^{'}}\in \mathcal{F}_{sk}.\hspace{3mm}\forall a,\nonumber \\
&\hspace{5mm}	\tilde{\sigma}^{B^{'}}\ge 0.\nonumber
\end{align}

Then the dual problem of $\mathrm{R}_{sc}(\mathsf{\Lambda})$ can be written as 
\begin{align}
1+	\mathrm{R}_{sc}(\mathsf{\Lambda})=&\hspace{5mm}\max \sum_a\mathrm{tr}(Y_aJ_a)\label{r7}\\
	\textit{s. t.}&\hspace{5mm} M-Y_a\in \mathcal{F}_{sk}^{*},\nonumber\\
	&\hspace{5mm}\mathrm{tr}_{V}M=\mathbb{I},\nonumber\\
	&\hspace{5mm}Y_a\ge 0.\nonumber
\end{align}

Next we show that $\mathrm{R}_{sc}(\cdot)$ satisfies the following properties:
\begin{itemize}
	\item[1.] [Faithfulness]$\Longleftarrow:$ If $\mathrm{R}_{sc}(\mathsf{\Lambda})=0$, then  due to the definition of $\mathrm{R}_{kDM}(\cdot)$, we have $\mathsf{\Lambda}\in \mathcal{F}_{sc}.$
	
	$\Longrightarrow:$ If $\mathsf{\Lambda}\in \mathcal{F}_{sc},$ when taking $\Gamma_i$ in (\ref{r5}) as $\Lambda_i$, $r=0$. As $r$ is nonnegative, $r=0$ is the minimum.
	
	\item[2.] [Convexity] Assume $\mathsf{\Lambda}_1=\{\Lambda_a^{(1)}\}$ and $\mathsf{\Lambda}_2=\{\Lambda_a^{(2)}\}$ are two sets of subchannels generated under the Definition \ref{d2}, $\{O_a\}$ and $\{Q_a\}$ are the optimal for $\mathsf{\Lambda}_1$ and $\mathsf{\Lambda}_2$ in terms of (\ref{r6}), respectively, then based on $\mathrm{R}_{sc}(\mathsf{\Lambda})$ defined in (\ref{r6}), $\mathrm{R}_{sc}(\mathsf{\Lambda}_1)=\sum_a\mathrm{tr}(O_a),$ $\mathrm{R}_{sc}(\mathsf{\Lambda}_2)=\sum_a\mathrm{tr}(Q_a)$. Let $J_a$ and $K_a$ be the Choi-Jamiolkowski isomorphism of $\Lambda^{(1)}_a$ and $\Lambda^{(2)}_a,$ respectively. Let $\mathsf{\Gamma}=p\mathsf{\Lambda}_1+(1-p)\mathsf{\Lambda}_2,$ $p\in (0,1),$
$pJ_a+(1-p)K_a$ are Choi-Jamiolkowski isomorphism of subchannels in $\mathsf{\Gamma},$
	\begin{align*}
		1+\mathrm{R}_{sc}(\Gamma)
		\le& p\sum_a\mathrm{tr}(O_a)+(1-p)\mathrm{tr}(Q_a)\\
		=&p[1+\mathrm{R}_{sc}(\mathsf{\Lambda_1})]+(1-p)[1+\mathrm{R}_{sc}(\mathsf{\Lambda_2})].
	\end{align*}
	
	The first inequality is due to that $pO_a+(1-p)Q_a\ge pJ_a+(1-p)K_a$. Hence, we finish the proof of $\mathrm{R}_{sc}(\Gamma)\le p\mathrm{R}_{sc}(\Lambda_1)+ (1-p)\mathrm{R}_{sc}(\Lambda_1).$
\end{itemize}
\vspace{5mm}
\subsection{Proof of Theorem \ref{t2}}\label{app2}

In this subsection, we mainly present the proof of Theorem \ref{t2}. Before proving Theorem \ref{t2}, we first present the dual problem of $\mathrm{R}_e(\cdot)$, then we get some Lemmas needed to show Theorem \ref{t2}.

Based on the knowledge of semidefinite programming given in Sec. \ref{app1}, we  present the dual problem of $\mathrm{R}_{ke}(\rho_{AB})$,
\begin{align}
	1+	\mathrm{R}_{ke}(\rho_{AB})=\hspace{5mm}&\max \mathrm{tr}(A\rho)\nonumber\\
	\textit{s. t.}\hspace{5mm}& \mathbb{I}-A\in \mathcal{F}_{sk}^{*},\nonumber\\
	&\hspace{5mm}A\ge0.\label{r8}
\end{align}

\begin{Lemma}
Let $\rho^{A^{'}B}$ be a bipartite mixed state, assume $\mathbb{M}^A=\{M_a^{AA^{'}}\}_a$ is an arbitrary POVM operated by Alice, $\mathsf{\Lambda}=\{\Lambda_i\}$ is a set of subchannels defined in (\ref{c}) generated by $\mathbb{M}^A$ and $\rho^{A^{'}B}$, then
\begin{align*}
	\mathrm{R}_{ke}(\rho)=\max_{\mathbb{M}^A}\mathrm{R}_{sc}(\mathsf{\Lambda}),
\end{align*}	
where the maximum takes over all the POVMs $\mathbb{M}^A.$
\end{Lemma}
\emph{Proof:} First we show that $r_1=\mathrm{R}_{ke}(\rho)\ge \mathrm{R}_{sc}(\Lambda).$ Let $\gamma_{A^{'}B}$ be the optimal for $\rho^{A^{'}B}$ in terms of $\mathrm{R}_{ke}(\rho_{A^{'}B})$, and let $\mathsf{\Gamma}=\{\Gamma_i(\cdot)\}$ be the set of subchannels such that
\begin{align*}
	\Gamma_i(\phi^A)=\mathrm{tr}_{AA^{'}}[(M_i^{AA^{'}}\otimes\mathbb{I}^B)(\phi^A\otimes\gamma^{A^{'}B})],
\end{align*}
 then we have 
\begin{align*}
	(1+r_1)\Gamma_i-\Lambda_i\in \mathcal{R}_{sc}.
\end{align*}

As $\gamma_{A^{'}B}$ is a state with $SN(\gamma)\le k,$ then $\mathsf{\Gamma}\in \mathcal{F}_{sc}.$ Due to the definition of $\mathrm{R}_{sc}(\Lambda)$, $\mathrm{R}_{ke}(\rho)\ge\mathrm{R}_{sc}(\Lambda).$

Next we present a teleportation instrument $\mathsf{\Lambda}$ such that $\mathrm{R}_{ke}(\rho)\le \mathrm{R}_{sc}(\Lambda).$ 
Assume $\{U_i\}$ is a set of Pauli operators with respect to the basis $\{\ket{i}^V\},$ let $A$ be the optimal for $\rho^{A^{'}B^{'}}$ in terms of $\mathrm{R}_{ke}(\rho^{A^{'}B})$. Let 
\begin{align*}
	Y^{VB}_a=&\frac{1}{d}(\mathcal{U}^V_a\otimes\mathbb{I})(A^{VB})^T,\\
	M^{VA^{'}}_a=&(\mathcal{U}^V_a\otimes\mathbb{I})(\psi_{+}),
\end{align*}
here $\mathcal{U}_a(\cdot)=U_a(\cdot)U_a^{\dagger}$, then 

\begin{widetext}
\begin{align*}
	 1+\max_{\mathbb{M}^A}\mathrm{R}_{sc}(\mathsf{\Lambda})\ge&\sum_a\mathrm{tr}Y_aJ_a\\
=&\sum_a\mathrm{tr}[(Y_a^{VB}\otimes\mathbb{I}^{AA^{'}})(M_a^{AA^{'}}\otimes\mathbb{I}^{VB})(\psi_{+}^{VA}\otimes\rho^{A^{'}B})]\\
	=&\sum_a\mathrm{tr}[(Y_a^{VB}\otimes\mathbb{I}^{AA^{'}})((M_a^{VA^{'}})^T\otimes\mathbb{I}^{VB})(\psi_{+}^{VA}\otimes\rho^{A^{'}B})]\\
	=&\frac{1}{d}\sum_a\mathrm{tr}[(Y_a^{VB}\otimes\mathbb{I}^{A^{'}})((M_a^{VA^{'}})^T\otimes\mathbb{I}^{B})(\mathbb{I}^{A}\otimes\rho^{A^{'}B})]\\
	=&\frac{1}{d^2}\sum_a\mathrm{tr}_{VA^{'}B}[(\mathcal{U}^V_a\otimes\mathbb{I})(A^{VB})^T\otimes\mathbb{I}^{A^{'}}][(\mathcal{U}^V_a\otimes\mathbb{I})(\psi_{+})](\mathbb{I}^V\otimes\rho^{A^{'}B})]\\
	=&\mathrm{tr}(A\rho)\\
	=&1+\mathrm{R}_{ke}(\rho^{A^{'}B}).
\end{align*}
\end{widetext}
Hence, we finish the proof.

\begin{Lemma}\label{la2}
	Assume $\rho^{AB}$ is a bipartite mixed state, $\mathbb{M}^A=\{M_i^{AA^{'}}\}_i$ is a POVM, $\mathsf{\Lambda}=\{\Lambda_i\}_i$ is a set of subchannels defined in (\ref{c}), then
	\begin{align*}
		\mathrm{R}_{sc}(\mathsf{\Lambda})=\max_{\mathbb{M}^B}(\mathbb{M}^{AB}),
	\end{align*}
	where $\mathbb{M}^{AB}$ is defined in (\ref{pm}), and the maximum takes over all the POVMs $\mathbb{M}^B=\{M^{BB^{'}}\}.$
\end{Lemma}
\emph{Proof:}
First we show that $\mathrm{R}_{sc}(\mathsf{\Lambda})$ is lower than $\max_{\mathbb{M}^B}(\mathbb{M}^{AB}).$

Let $\{Y_a\}$ and $M$ be the optimal for $\mathsf{\Lambda}$ in terms of (\ref{r7}), let $\{U_b^B\}$ be a set of Pauli operators with respect to the basis $\{\ket{i}_B\}$. Let 
\begin{align*}
	M_b^{BB^{'}}=&( \mathcal{U}_b\otimes\mathbb{I})(\psi_{+}^{{BB^{'}}})\\
	A_{ab}=&(\mathcal{U}_b\otimes\mathbb{I})(Y_a^{T}),\\
	 K=&\frac{1}{d}(\mathcal{U}_b\otimes\mathbb{I})(M^T),
\end{align*}
here $\mathcal{U}_b(\cdot)=U_b(\cdot)U_b^{\dagger}$.

Based on the definitions of $\{Y_a\}$ and $M$, $\{A_{ab}\}$ and $K$ satisfy the restricted conditions in $(\ref{r4})$, hence we have
\begin{widetext}
\begin{align}
	&1+\max_{\mathbb{M}^B} \mathrm{R}_{B}(\mathbb{M}^{AB})
\\\ge&\sum_{ab}\mathrm{tr}(A_{ab}M_{ab})\nonumber\\
	=&\sum_{ab}\mathrm{tr}[(\mathcal{U}_b\otimes\mathbb{I})(Y_a^{T})\otimes\mathbb{I}_{A^{'}B^{'}}][(M^{AA^{'}}_a\otimes (\mathcal{U}_b\otimes\mathbb{I})(\psi_{+}^{BB^{'}}))(\mathbb{I}^A\otimes\rho_{A^{'}B^{'}}\otimes\mathbb{I}^B)]\nonumber\\
	=&\frac{1}{d^2}\sum_{ab}\mathrm{tr}[Y_a^{AB^{'}}(M_a^{AA^{'}}\otimes \mathbb{I}^{BB^{'}})(\psi_{+}^{VA}\otimes\rho^{A^{'}B^{'}})]\\
	=&\sum_a\mathrm{tr}[Y_aJ_a]\nonumber\\
	=&1+\mathrm{R}_{sc}(\mathsf{\Lambda}).\label{f1}
\end{align}
\end{widetext}
Then we show the inequality of the other side. Assume $\mathbb{M}^{AB}=\{M_{ab}^{AB}\}$ is generated under the fomula (\ref{pm}), then we could construct $\mathbb{N}^{VB}=\{N_{ab}\}$ such that
\begin{align*}
	N_{ab}^{VB}=d\mathrm{tr}_{A}[(\mathbb{I}^V\otimes M_{ab}^{AB})(\psi_{+}^{VA}\otimes \mathbb{I}^B)],
\end{align*}

here $\psi_{+}^{VA}$ is the maximally entangled state. As $(\mathbb{I}\otimes A)\ket{\psi}=(A^T\otimes\mathbb{I})\ket{\psi}$, $N_{ab}^T=M_{ab}$.

\begin{align}
	&N_{ab}^{VB}\\
	=&d\mathrm{tr}_{AA^{'}B^{'}}[(\mathbb{I}^V\otimes M_a^{AA^{'}}\otimes M^{BB^{'}}_b)(\psi_{+}^{VA}\otimes\rho^{A^{'}B^{'}}\otimes\mathbb{I}^B) ]\\
	=&\mathrm{tr}_{B^{'}}[(\mathbb{I}^V\otimes M_b^{BB^{'}})(J_a^{VB^{'}}\otimes\mathbb{I}^B)],\label{f2}
\end{align}
the second equality is due to the definition of $\Lambda_i(\cdot)$ defined in Definition \ref{d2}, the Choi-Jamiolkowski isomorphism of subchannels $\Lambda_a$, and $(\mathbb{I}\otimes A)\ket{\psi}=(A^T\otimes\mathbb{I})\ket{\psi}$. Due to the definition of $\mathrm{R}_{sc}(\mathsf{\Lambda})$, we have
\begin{align*}
	J_a^{VB^{'}}\le [1+\mathrm{R}_{sc}(\mathsf{\Lambda})]O_a,\hspace{3mm} \{O_a\}\in \mathcal{F}_{sk},
\end{align*}

combing the above formula, (\ref{f2}) can be turned into 
\begin{align}
	N_{ab}^{VB}\le&[1+\mathrm{R}_{sc}(\mathsf{\Lambda})]\mathrm{tr}_{B^{'}}[(\mathbb{I}^V\otimes M_b^{BB^{'}})(O_a^{VB^{'}}\otimes\mathbb{I}^B)]\nonumber\\
	=&[1+\mathrm{R}_{sc}(\mathsf{\Lambda})]T_{ab},\label{r10}
\end{align}
here $\{T_{ab}\}$ is in $\mathcal{F}_{B}.$ Then we have 
\begin{align*}
	N^{AB}_{ab}\le [1+\mathrm{R}_{sc}(\mathsf{\Lambda})]T_{ab}^{AB}.
\end{align*}

Hence, due to the definition of $\mathrm{R}_{kDM}(\mathbb{M}^{AB})$, and the set $\mathcal{F}_{kDM}$ is closed under the transpose $(\cdot)^T$, we have $\max_{\mathbb{M}^B}\mathrm{R}_{kDM}(\mathbb{M}^{AB})\le \mathrm{R}(\mathsf{\Lambda})$.

 At last, by combing (\ref{f1}), we have $\max_{\mathbb{M}^B}\mathrm{R}_{kDM}(\mathbb{M}^{AB})\le \mathrm{R}(\mathsf{\Lambda}).$ At last, we show $\max_{\mathbb{M}^B}\mathrm{R}_{kDM}(\mathbb{M}^{AB})= \mathrm{R}(\mathsf{\Lambda}).$

\vspace{2mm}

\emph{	Proof of Theorem \ref{t2}:}
First we prove $\mathrm{R}_{ke}(\rho_{AB})\ge\mathrm{R}_{kDM}(\mathbb{M}).$ Assume $\mathrm{R}_{ke}(\rho_{AB})=r,$ $\sigma_{AB}$ and $\gamma_{AB}$ are the states defined in (\ref{r0}) for $\mathrm{R}_{ke}(\rho_{AB})$, then let 
\begin{align*}
	M^{'}_{ab}=&\mathrm{tr}_{A^{'}B^{'}}[(M_a^{AA^{'}}\otimes M_b^{BB^{'}})(\mathbb{I}_A\otimes\sigma_{A^{'}B^{'}}\otimes\mathbb{I}_B)],\\
		M_{ab}^{''}=&\mathrm{tr}_{A^{'}B^{'}}[(M_a^{AA^{'}}\otimes M_b^{BB^{'}})(\mathbb{I}_A\otimes\gamma_{A^{'}B^{'}}\otimes\mathbb{I}_B)].
\end{align*}
Due to the above definition, we have $\{M_{ab}^{''}\}\in \mathcal{F}_{kDM},$ and $M_{ab}\le (1+r)O_{ab}$. By the definition of $\mathrm{R}_{kDM}(\cdot),$ we finish the proof of $\mathrm{R}_e(\rho_{AB})\ge\mathrm{R}_{kDM}(\mathbb{M}).$

At last, we show the other side of the inequality. By Lemma \ref{la2}, we have

\begin{align*}
	\max_{\mathbb{M}^A,\mathbb{M}^B}\mathrm{R}_{kDM}(\mathbb{M}^{AB})=\max_{\mathbb{M}^A}[\max_{\mathbb{M}^B}\mathrm{R}_{kDM}(\mathbb{M}^{AB})]=\max_{\mathbb{M}^A}\mathrm{R}(\mathsf{\Lambda}^{A\rightarrow B^{'}}).
\end{align*}
That is, if we prove $\max_{\mathbb{M}^A}\mathrm{R}(\mathsf{\Lambda}^{A\rightarrow B^{'}})\ge\mathrm{R}_e(\rho_{AB})$, we finish the proof.

Assume $A$ is the optimal for $\rho$ in terms of $\mathrm{R}_{ke}(\cdot)$. Let $\{U_a\}_{a=1}^{d^{'}}$ be a set of Pauli operators with respect to $\{\ket{i}^{A^{'}}\}$. Let $\mathbb{M}^A=\{M_a^{AA^{'}}=(\mathbb{I}\otimes \mathcal{U}_a)[\psi_{+}]\}$. Based on the above assumption, let $M$, $Y_a$ and $\mathbb{M}^A$ be the following
\begin{align*}
	M=\frac{1}{d}\mathbb{I},\hspace{3mm}Y_a=(\mathbb{I}\otimes \mathcal{U}_a)[A],\hspace{4mm}M_a^{AA^{'}}=(\mathbb{I}\otimes\mathcal{U}_a)(\psi_{+}),
\end{align*}which are appropriate for $\mathrm{R}_{sc}(\mathsf{\Lambda})$ in terms of  (\ref{r7}).

Hence, we have
\begin{align}
&	1+\max_{\mathbb{M}^A}\mathrm{R}(\mathsf{\Lambda}^{A\rightarrow B^{'}})\nonumber\\
	\ge&\sum_a\mathrm{tr}(Y_aJ_a)\nonumber\\
	=&\sum_a\mathrm{tr}[(\mathbb{I}\otimes\mathcal{U}_a^{}[A])\mathrm{tr}_{AA^{'}}[(\mathbb{I}^V\otimes M_a^{AA^{'}}\otimes\mathbb{I}^{B^{'}})(\psi_{+}\otimes\rho) ]]\nonumber\\
	=&\sum_a\mathrm{tr}[(\mathbb{I}\otimes\mathcal{U}_a[A])\otimes\mathbb{I}_{AA^{'}}][(\mathbb{I}^V\otimes (\mathbb{I}\otimes\mathcal{U}_a)(\psi_{+})\otimes\mathbb{I}^{B^{'}})(\psi_{+}\otimes\rho) ]\nonumber\\
	=&\mathrm{tr}(A\rho)\nonumber\\
	=&1+\mathrm{R}_{ke}(\rho).\label{r11}
\end{align}
Then we finish the proof of this theorem.

\subsection{Proof of Proposition \ref{t5}}\label{app3}
 In the proof, let $\mathrm{R}_{ke}(\rho)=r,$ and $\gamma$ is the optimal in terms of $\mathrm{R}_{ke}(\cdot)$ for $\rho$, 
 \begin{align*}
 	\rho_{AB}\le (1+r)\gamma_{AB}, \hspace{4mm} SN(\gamma_{AB})\le k.
 \end{align*} 
Then
\begin{widetext}
\begin{align*}
p_{g}(\mathcal{G},\rho_{AB})=&\max_{\mathbb{M}^A,\mathbb{M^B}}\sum_{a,b,x,y}p_{xy}\mathrm{tr}(M_{ab}\sigma_{xy})\\
=&\max_{\mathbb{M}^A,\mathbb{M^B}}\sum_{a,b,x,y}p_{xy}\mathrm{tr}(M_x^{AA^{'}}\otimes M_y^{BB^{'}})(\rho^{AB}\otimes\mathbb{I}^{A^{'}B^{'}})(\mathbb{I}^{AB}\otimes\sigma^{A^{'}B^{'}}_{xy})\\
\le& (1+r)\max_{\mathbb{M}^A,\mathbb{M^B}}\sum_{a,b,x,y}p_{xy}\mathrm{tr}(M_x^{AA^{'}}\otimes M_y^{BB^{'}})(\gamma^{AB}\otimes\mathbb{I}^{A^{'}B^{'}})(\mathbb{I}^{AB}\otimes\sigma^{A^{'}B^{'}}_{xy}),\\
	\frac{p_g(\mathcal{G},\rho_{AB})}{p_g^{(k)}(\mathcal{G})}=&\frac{\sum_{a,b,x,y}p_{xy}\mathrm{tr}(M_x^{AA^{'}}\otimes M_y^{BB^{'}})(\rho^{AB}\otimes\mathbb{I}^{A^{'}B^{'}})(\mathbb{I}^{AB}\otimes\sigma^{A^{'}B^{'}}_{xy})}{\max_{\delta\in \mathcal{F}_{sk}}\sum_{a,b,x,y}p_{xy}\mathrm{tr}(M_x^{AA^{'}}\otimes M_y^{BB^{'}})(\delta^{AB}\otimes\mathbb{I}^{A^{'}B^{'}})(\mathbb{I}^{AB}\otimes\sigma^{A^{'}B^{'}}_{xy})}\\
	\le& \frac{(1+r)\sum_{a,b,x,y}p_{xy}\mathrm{tr}(M_x^{AA^{'}}\otimes M_y^{BB^{'}})(\delta^{AB}\otimes\mathbb{I}^{A^{'}B^{'}})(\mathbb{I}^{AB}\otimes\sigma^{A^{'}B^{'}}_{xy})}{\max_{\delta\in \mathcal{F}_{sk}}\sum_{a,b,x,y}p_{xy}\mathrm{tr}(M_x^{AA^{'}}\otimes M_y^{BB^{'}})(\delta^{AB}\otimes\mathbb{I}^{A^{'}B^{'}})(\mathbb{I}^{AB}\otimes\sigma^{A^{'}B^{'}}_{xy})}\\
	\le&1+r.
\end{align*}
\end{widetext}
The first inequality is due to that $\delta$ is the optimal in terms of $\mathrm{R}_{ke}(\cdot)$ for $\rho$. Hence, we have
\begin{align*}
	\max_{\mathcal{G}}\frac{p_g(\mathcal{G},\rho_{AB})}{p_g^{(k)}(\mathcal{G})}\le 1+\mathrm{R}_{ke}(\rho_{AB}).
\end{align*}

Next we prove the other side of the inequality, $	\max_{\mathcal{G}}\frac{p_g(\mathcal{G},\rho_{AB})}{p_g^{(k)}(\mathcal{G})}\ge 1+\mathrm{R}_{ke}(\rho_{AB}).$ Assume $F$ is the optimal for $\rho$ in terms of $\mathrm{R}_{ke}(\rho_{AB})$, let $\tilde{F}=\frac{F}{\mathrm{tr}F},$ and $p_{kl}=\frac{1}{d^4}.$ Let $\psi_{+}$ be the maximally entangled state, $\mathbb{M}^{A}=\{\mathbb{I}\otimes\mathcal{U}^{A^{'}}_k(\psi^{AA^{'}}_{+})\},$ $\mathbb{M}^B=\{\mathbb{I}\otimes\mathcal{U}^{B^{'}}_l(\psi_{+}^{BB^{'}})\}$, $\sigma_{kl}=\mathbb{I}^{AB}\otimes\mathcal{U}_k^{A^{'}}\otimes\mathcal{U}_l^{B^{'}}(\tilde{F}),$ here $\psi_{+}$ is the maximally entangled state, then 

\begin{align}
&	p_{g}(\mathcal{G},\rho_{AB})\nonumber\\=&\sum_{k,l}\frac{1}{d^4}\mathrm{tr}(M_k^{AA^{'}}\otimes M_l^{BB^{'}})(\rho^{AB}\otimes\mathbb{I}^{A^{'}B^{'}})(\mathbb{I}^{AB}\otimes\sigma^{A^{'}B^{'}}_{kl})\nonumber\\
	=&\sum_{k,l}\frac{1}{d^4}\mathrm{tr}(\psi_{+}^{AA^{'}}\otimes\psi_{+}^{BB^{'}})(\rho^{AB}\otimes\mathbb{I}^{A^{'}B^{'}})(\mathbb{I}^{AB}\otimes \tilde{F})\nonumber\\
	=&\mathrm{tr}\frac{\rho F}{\mathrm{tr}F},\label{l1}
\end{align}
assume $\delta^{AB}$ is an arbitrary bipartite state with $SN(\delta^{AB})\le k,$
\begin{align}
	p_{g}(\mathcal{G},\delta^{AB})=
	\mathrm{tr}\frac{\delta F}{\mathrm{tr}F}\le \frac{1}{\mathrm{tr}F},\label{l2}
\end{align}
then combing $(\ref{l1})$ and $(\ref{l2}),$ 
\begin{align*}
	\frac{p_{succ}(\mathcal{G},\rho^{AB})}{\max_{\delta\in\mathcal{F}_{sk}}p_{succ}(\mathcal{G},\delta^{AB})}\ge& \frac{\mathrm{tr}\frac{\rho F}{\mathrm{tr}F}}{\frac{1}{\mathrm{tr}F}}\nonumber\\
	=&\mathrm{tr}(\rho F)\nonumber\\
	=&\mathrm{R}_e(\rho)+1.
\end{align*}
Hence, we finish the proof.

\bibliographystyle{IEEEtran}
\bibliography{ref}
\end{document}